\documentclass[aps,twocolumn,preprintnumbers,superscriptaddress,prl,10pt]{revtex4-1}
\usepackage{bbm}
\usepackage{bm}
\usepackage{mathrsfs }
\usepackage{amsmath}
\usepackage{amssymb}
\usepackage{empheq}
\usepackage{graphicx}
\usepackage{mathrsfs}
\usepackage{amsfonts}
\usepackage{amsthm}
\usepackage{color}
\usepackage{multirow}
\usepackage{bigints}
\usepackage{txfonts}
\usepackage{float}
\usepackage{hyperref}
\hypersetup{
     unicode=false,              
     pdftoolbar=true,           
     pdfmenubar=true,         
     pdffitwindow=false,      
     pdfstartview={FitH},    
     pdftitle={My title},       
     pdfauthor={Author},     
     pdfsubject={Subject},   
     pdfcreator={Creator},   
     pdfproducer={Producer}, 
     pdfkeywords={keyword1} {key2} {key3}, 
     pdfnewwindow=true, 
     colorlinks=false,       
     linkcolor=red,           
     citecolor=green,        
     filecolor=magenta,    
     urlcolor=cyan           
}

\makeatletter \tolerance = 10000 \tolerance = 10000

\makeatother
\begin{document}

\title{Skin-Anderson transitions in disordered hybrid-nonreciprocal systems}

\author{C. Wang}
\email[Corresponding author: ]{physcwang@tju.edu.cn}
\affiliation{Center for Joint Quantum Studies and Tianjin Key Laboratory of Low Dimensional Materials Physics and Preparing Technology, Department of Physics, School of Science, Tianjin University, Tianjin 300350, China}
\author{X. R. Wang}
\affiliation{School of Science and Engineering, Chinese University of Hong Kong (Shenzhen), Shenzhen 518172, China}
\author{Hechen Ren}
\affiliation{Center for Joint Quantum Studies and Tianjin Key Laboratory of Low Dimensional Materials Physics and Preparing Technology, Department of Physics, School of Science, Tianjin University, Tianjin 300350, China}
\affiliation{Joint School of National University of Singapore and Tianjin University, International Campus of Tianjin University, Binhai New City, Fuzhou 350207, China}

\date{\today}

\begin{abstract}
Anderson localization is a universal wave phenomenon characterized by a disorder-induced quantum phase transition from extended to localized states, while the skin effect is a universal phenomenon in non-Hermitian systems that allows chiral boundary states. Here, we report an unexpected skin-Anderson transition arising from the interplay between Anderson localization and the non-Hermitian skin effect in hybrid-nonreciprocal systems that exhibit both reciprocity and nonreciprocity in different spatial directions. In weakly disordered regimes, the states are boundary-extended, meaning they are spatially extended in one dimension but localized at the boundaries in others. As the disorder increases, these boundary-extended states transition to boundary-localized states at a critical disorder strength. Under a similar transformation, these critical points exhibit universal characteristics akin to those of the Anderson transitions in the corresponding Hermitian systems, including identical critical exponents within numerical errors. As disorder increases beyond a second critical threshold, a further transition occurs where boundary-localized states become bulk-localized, leading to the disappearance of the skin effect. These skin-Anderson transitions, which merge Anderson localization with the non-Hermitian skin effect, establish a new framework for controlling the localization properties of states in disordered systems.
\end{abstract}

\maketitle

\emph{Introduction.}$-$When disorder in a disordered medium becomes sufficiently strong, states within that medium will typically become localized to a specific region~\cite{pwAnderson_pr_1958}. The conventional scaling theory suggests that states in weakly disordered Hermitian systems can propagate freely under certain conditions of symmetry and spatial dimension~\cite{eAbrahams_prl_1979,paLee_rmp_1985,bKramer_rpp_1993,bHuckestein_rmp_1995,fEvers_rmp_2008}. As disorder increases, these extended states transition to localized states, a phenomenon known as the Anderson (localization) transition~\cite{dfriedan_prl_1980,shimami_prp_1980,ammpruisken_prl_1988,snevangelou_prl_1995,yasada_prl_2002,cwang_prl_2015,jhpixley_prl_2015,czChen_prl_2015,bFu_prl_2017,zXiao_prl_2023}. On the other hand, non-Hermiticity introduces a distinct localization mechanism, leading to different localization behaviors under periodic versus open boundary conditions~\cite{nHatano_prl_1996,yap_prl_2018,vmMartinez_prb_2018,rHamazaki_prl_2019,ljZhai_prb_2020,kSuthar_prb_2022}. This phenomenon is particularly evident in nonreciprocal systems with asymmetric hopping, where states are chiral and localize at certain boundaries, known as the non-Hermitian skin effect~\cite{yap_prl_2018}. Such boundary localization stems from the interplay of nonreciprocity and topology, manifesting as a specific bulk-boundary correspondence~\cite{teLee_prl_2016,dLeykam_prl_2017,hSun_prl_2018,zGong_prx_2018,dsBorgnia_prl_2020}.
\par

\begin{figure}[htbp]
\includegraphics[width=0.48\textwidth]{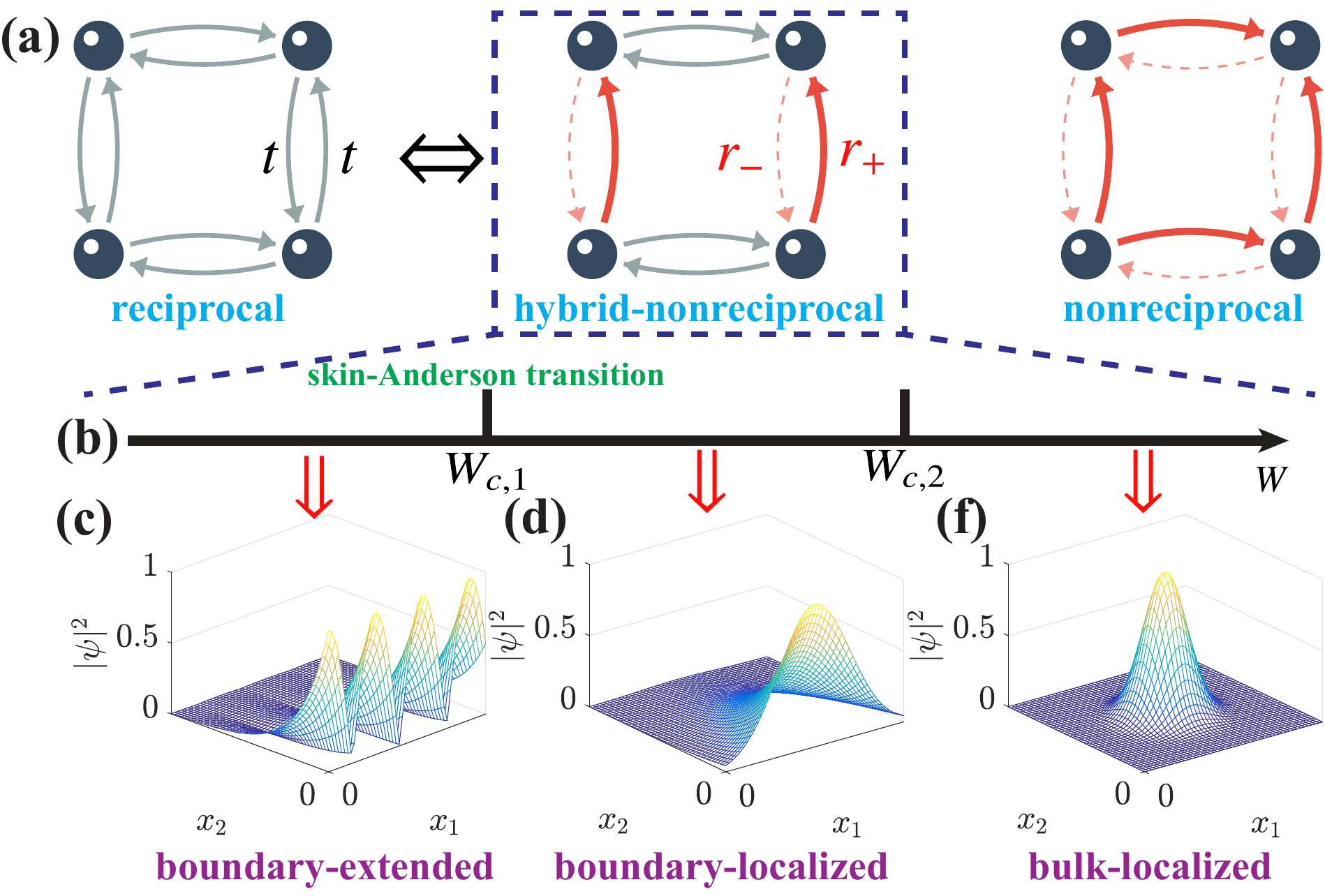}\centering
\caption{(a) A schematic diagram illustrates the reciprocal, hybrid-nonreciprocal, and nonreciprocal systems in two dimensions (2D). The hybrid-nonreciprocal system exhibits both symmetric (denoted as \( t \)) and asymmetric (denoted as \(r_{\pm}\)) hopping, whose transfer matrices can be mapped to its reciprocal counterpart under certain conditions. (b) A generalized phase diagram for hybrid-nonreciprocal systems indicates two critical transitions: a skin-Anderson transition and a boundary-to-bulk localization transition at \(W_{c,1}\) and \(W_{c,2}\), respectively. (c), (d) and (f) depict schematic representations of boundary-extended, boundary-localized, and bulk-localized states, three typical phases in disordered hybrid-nonreciprocal systems in 2D.}
\label{fig1}
\end{figure}

While much existing work has investigated the two localization mechanisms separately~\cite{yHuang_prb_2020,cWang_prb_2020,xLuo_prl_2021,kKawabata_prl_2021,xLuo_prresearch_2022,cWang_prb_2022,cWang_prb_2023,cWang_prb_2024,bLi_prl_2025,jShang_prb_2025}, the influence of non-Hermitian skin effects on the Anderson transitions remains poorly understood. Here, we report a new type of localization transition in hybrid-nonreciprocal systems that exhibit both disorder and the non-Hermitian skin effect, in which Anderson-type criticality in the reciprocal limit splits into two critical points, thereby separating three distinct phases. In weakly disordered states, wave functions exhibit both extended and localized characteristics: they decay exponentially along some spatial dimensions but spread uniformly across the entire sample along others. In mediated disorders, such a boundary-extended state becomes boundary-localized, with wave functions concentrating near the system boundary and losing spatial extension in the other directions. Finally, it transitions to a bulk-localized state as the disorder increases. These features are summarized in Fig.~\ref{fig1}.
\par

We term these transitions, from boundary-extended to boundary-localized states, as skin-Anderson transitions. Different from standard Anderson transitions~\cite{bKramer_rpp_1993,bHuckestein_rmp_1995,fEvers_rmp_2008}, which describe the shift from bulk-extended to bulk-localized states, skin-Anderson transitions are confined to boundary modes. Remarkably, we prove that the transfer matrices of hybrid-nonreciprocal systems can be mapped to that of its reciprocal counterpart through a similar transformation such that the critical behavior of skin-Anderson transitions closely resembles that of Anderson transitions: The correlation length \( \xi \) diverges at the critical disorder \( W_{c} \) as \( \xi \propto |W - W_c|^{-\nu} \) with \( \nu \) being the critical exponent and \( W \) measuring the strength of disorders. Consequently, skin-Anderson transitions provide new insights into the manipulation of extended and localized states in disordered media.
\par

\emph{Hybrid-nonreciprocal systems.}$-$Skin-Anderson transitions happen at hybrid-nonreciprocal systems, which are neither like reciprocal systems with symmetric hopping, nor like the generalized Hatano-Nelson model, where hopping terms in all spatial directions are asymmetric~\cite{nHatano_prb_1998}. Hybrid-nonreciprocal systems combine both reciprocity and nonreciprocity, in which at least one spatial dimension is reciprocal; see Fig.~\ref{fig1}(a) for a schematic plot. Hereafter, we apply open boundary conditions to introduce the non-Hermitian skin effect.  
\par

For simplicity, we assume that each spatial direction has the same degree of nonreciprocity and stipulate that there is only one reciprocal direction (e.g., the \(x_1\)-direction). Then, a general tight-binding Hamiltonian (with unit lattice constant) for hybrid-nonreciprocal systems reads as
\begin{equation}
    \begin{gathered}
        H=\sum_{\bm{i}}\left\{ c^\dagger_{\bm{i}}\epsilon_{\bm{i}}
        +\sum_{\mu}\left[ c^\dagger_{\bm{i}+\hat{x}_{\mu}}r_+V_{\bm{i}+\hat{x}_\mu,\bm{i}}
        +c^\dagger_{\bm{i}-\hat{x}_{\mu}}r_-V_{\bm{i}-\hat{x}_\mu,\bm{i}} \right] \right\}c_{\bm{i}}
    \end{gathered}\label{eq_1}
\end{equation}
with \( c^\dagger_{\bm{i}} \) and \( c_{\bm{i}} \) being the single-particle creation and annihilation operators on a lattice site \( \bm{i}=(x_1,x_2,\cdots,x_d) \). \( \hat{x}_{\mu} \) represents the unit vector of the \(\mu-\)th spatial direction. The on-site energy $\epsilon_{\bm{i}}$ is taken to be a Hermitian matrix ($\epsilon_{\bm{i}}=\epsilon^\dagger_{\bm{i}}$), and the hopping matrices satisfy $V_{\bm{i}+\hat{x}_\mu,\bm{i}}=V^\dagger_{\bm{i},\bm{i}+\hat{x}_{\mu}}$. The nonreciprocity is encoded in the coefficients \( r_{\pm}=1\pm \Tilde{t}/t \) with \( t \) and \( \tilde{t} \) being real positive numbers and measuring the strengths symmetric and asymmetric hopping, respectively. We set \( r_+=r_-=1 \) for \( \mu=1 \) to ensure reciprocity along the \(x_1-\)direction.
\par

We can classify hybrid-nonreciprocal systems into three categories: under-nonreciprocal (\( \tilde{t}<t \)), critically nonreciprocal (\( \tilde{t}=t \)), and over-nonreciprocal (\( \tilde{t}>t \)). In the under-nonreciprocal regime (\( \tilde{t}<t \)), we show that the system's energy spectrum is always real; see the proof in the Supplementary~\cite{supp}. This conclusion can be understood through the following reasoning: For under-nonreciprocity \( \tilde{t}<t \), one can find an invertible matrix \( S \) that transforms the Hamiltonian of the hybrid-nonreciprocal system into a Hermitian Hamiltonian via a similarity transformation, i.e., \( \tilde{H} = SHS^{-1} \)~\cite{supp}. For model~\eqref{eq_1}, we can choose the invertible matrix as  
\begin{equation}
    \begin{gathered}
        S=\sum_{\bm{i}}c^\dagger_{\bm{i}} \left[\prod^d_{\mu=2}\left(\dfrac{t-\tilde{t}}{t+\tilde{t}} \right)^{x_{\mu}/2}\right] c_{\bm{i}}
    \end{gathered}\label{eq_2}
\end{equation}
such that
\begin{equation}
    \begin{gathered}
        \tilde{H}=\sum_{\bm{i}} \left[ c^\dagger_{\bm{i}}\epsilon_{\bm{i}}c_{\bm{i}}+\sum_{\bm{\mu}}\dfrac{1}{\gamma_\mu}\left( c^\dagger_{\bm{i}+\hat{x}} V_{\bm{i}+\hat{x}_{\mu},\bm{i}} c_{\bm{i}}+h.c. \right) \right].
    \end{gathered}\label{eq_3}
\end{equation}
In Eq.~\eqref{eq_3}, \( \gamma_\mu = 1\) for \( \mu=1 \) (the reciprocal direction), whereas \( \gamma_\mu = (1-\tilde{t}^2/t^2)^{-1/2} \) for all other cases. Since \( \tilde{H} \) is Hermitian, the corresponding matrix \( H \) should also have a real energy spectrum.
\par

However, the Hamiltonian of the hybrid-nonreciprocal system cannot be mapped into a Hermitian matrix via the above transformation for \( \tilde{t}\geq t \)~\cite{supp}. In the over-nonreciprocal case (\( t<\tilde{t} \)), the system's energy spectrum is always complex. Critical nonreciprocity (\( t=\tilde{t} \)) marks the transition between the under- and over-nonreciprocal regimes, where the system exhibits exceptional points. Next, we focus on the under-nonreciprocal case (\( t>\tilde{t} \)), where skin-Anderson transitions happen.
\par

\emph{Skin-Anderson transition.}$-$Now, we study the localization problems in hybrid-nonreciprocal systems within the framework of the transfer-matrix method~\cite{cwang_prl_2015}. We consider a quasi-\(d-\)dimensional long strip along the \(x_1-\)direction (reciprocity along this direction). A transfer matrix \( T_n \) reads as~\cite{supp}
\begin{equation}
    \begin{gathered}
        \begin{bmatrix}
            \psi_{n+1} \\
            U_{n+1,n}\psi_n
        \end{bmatrix}=
        \underbrace{
        \begin{bmatrix}
            U^{-1}_{n,n+1}(E-H_n) & -U^{-1}_{n.n+1} \\
            U_{n+1,n} & 0
        \end{bmatrix}}_{= T_n}
        \begin{bmatrix}
            \psi_{n} \\
            U_{n,n-1}\psi_{n-1}
        \end{bmatrix}
    \end{gathered}\label{eq_4}
\end{equation}
with \( E \) being the energy (real for the under-nonreciprocity) and \( \psi_n \) representing the wave functions of the \( n \)th slide of the long strip. \( U_{n,m} \) is the connection matrix between the \( n \)th and \( m \)th slide, which can be written as \( U_{n,m}=\sum_{\bm{i}'}c^\dagger_{\bm{i}'}V_{n\oplus \bm{i}',m\oplus\bm{i}'}c_{\bm{i}'} \). Since the long strip is along the \(x_1\)-direction, which is reciprocal, one can prove \( U^\dagger_{n,m}=U_{n,m} \). \( H_n \) is the Hamiltonian of the \( n \)th, which are non-Hermitian and nonreciprocal for a hybrid-nonreciprocal system, i.e., \( H_n\neq H^\dagger_n \). 
\par

For the strip of length \( L \), the total transfer matrix is expressed as \( T = \prod_{n=1}^{L} T_n \). In Hermitian (reciprocal) systems, the total transfer matrix \( T \) is symplectic such that the eigenvalues of the matrix \( \Omega = \lim_{L \to \infty} \ln{[ T T^\dagger ]}/(2L) \) appear in symmetric pairs, i.e., \( \gamma_1 > \gamma_2 > \ldots > \gamma_M > 0 > -\gamma_M > \ldots > -\gamma_1 \). Here, \( \gamma_{1,2,\ldots,M} \) are referred to as Lyapunov exponents~\cite{aCrisanti_springer_1993}. The finite-size localization length \( \lambda_M \), which characterizes the localization behavior of disordered systems, is given by the inverse of the smallest positive Lyapunov exponent, i.e., \( \lambda_M = 1/\gamma_M \). However, when extending this concept to non-Hermitian Hamiltonians, complications can arise, as the total transfer matrix \( T \) may be non-symplectic, making it difficult to identify the smallest positive Lyapunov exponent accurately~\cite{rkKunst_prb_2019,xLuo_prb_2021,yFu_prb_2023}.
\par

Notably, we prove that the Lyapunov exponents of hybrid-nonreciprocal systems also appear in symmetric pairs, so the previously established method for defining localization lengths remains valid~\cite{supp}. This characteristic can be demonstrated through the transformation of the wave function, expressed as \( \tilde{\psi}_{n} = S\psi_n \), with \( S \) given by Eq.~\eqref{eq_2}. Effectively, the transfer matrix is under a similar transformation:~\cite{supp}
\begin{equation}
    \begin{gathered}
    \tilde{T}_n=\begin{bmatrix}
        S & 0 \\
        0 & S
    \end{bmatrix}T_n
    \begin{bmatrix}
        S & 0 \\
        0 & S
    \end{bmatrix}^{-1}=
        \begin{bmatrix}
            U^{-1}_{n,n+1}(E-\tilde{H}_n) & -U^{-1}_{n.n+1} \\
            U_{n+1,n} & 0
        \end{bmatrix}
    \end{gathered}\label{eq_6}
\end{equation}
with \( \tilde{H}_n \) being given by Eq.~\eqref{eq_3}.
\par

It is easy to see that \( \tilde{T}_n \) in Eq.~\eqref{eq_6} stands for the transfer matrix of the reciprocal counterpart of the hybrid-nonreciprocal Hamiltonian~\eqref{eq_1}, i.e., model~\eqref{eq_3}. Given that similarity transformations preserve eigenvalues (and thus Lyapunov exponents), we conclude that the localization length of the hybrid-nonreciprocal system (Hamiltonian~\eqref{eq_1}) along its reciprocal direction is identical to that of its Hermitian counterpart (Hamiltonian~\eqref{eq_3}). Consequently, should the Hermitian counterpart undergo an Anderson transition from extended to localized states, the hybrid-nonreciprocal model~\eqref{eq_1} will exhibit an analogous localization transition along the \( x_1 \)-direction, characterized by the same critical exponents.
\par

Nevertheless, this phenomenon does not represent an Anderson transition. The underlying reason lies in the inverse transformation required to reconstruct the wave functions of the hybrid-nonreciprocal system, i.e., \( \psi_n=S^{-1}\tilde{\psi}_n \), which imparts a universal skin effect along the nonreciprocal directions~\cite{yap_prl_2018}, and we quantify its associated skin depth \( \delta = 2/|\ln(|t+\tilde{t}|/|t-\tilde{t}|)| \)~\cite{supp}. In essence, the extended and localized states of the Hermitian system are converted into states that inherently possess a non-Hermitian skin effect. We therefore categorize this localization transition within the hybrid-nonreciprocal system as a skin-Anderson transition.
\par

\emph{Hybrid-nonreciprocal Rashba model.}$-$One of the platforms for realizing the skin-Anderson transition is the hybrid-nonreciprocal Rashba model on a 2D square lattice described by Eq.~\eqref{eq_1} with
\begin{equation}
    \begin{gathered}
        \epsilon_{\bm{i}}=u_{\bm{i}}\sigma_0, V_{\bm{i},\bm{i}+\hat{x}_1}=t\sigma_0-i\alpha\sigma_1, V_{\bm{i},\bm{i}+\hat{x}_2}=t\sigma_0+i\alpha\sigma_2.
    \end{gathered}\label{eq_7}
\end{equation}
Here, \( u_{\bm{i}} \) is a white noise and is distributed uniformly in the range of \( [-W/2, W/2] \). \( \alpha \) is a real positive number that measures the strength of a Rashba-like (pseudo-)spin-orbit interaction~\cite{cWang_prb_2017,rSepehrinia_prb_2010}. \( \sigma_0 \) and \( \{ \sigma_{\mu=1,2,3} \} \) stand for the unit and Pauli matrices, respectively. In the reciprocal limit, the Rashba model belongs to the Gaussian symplectic ensemble and allows Anderson transitions~\cite{fEvers_rmp_2008,aAltland_prb_1997}. Therefore, the hybrid-nonreciprocal Rashba model should undergo a skin-Anderson transition, as shown in the above analysis. 
\par

We determine the critical disorder \( W_{c,1} \) and exponent \( \nu \) of skin-Anderson transitions through two independent numerical approaches.
\par

\begin{figure}[htbp]
\includegraphics[width=0.48\textwidth]{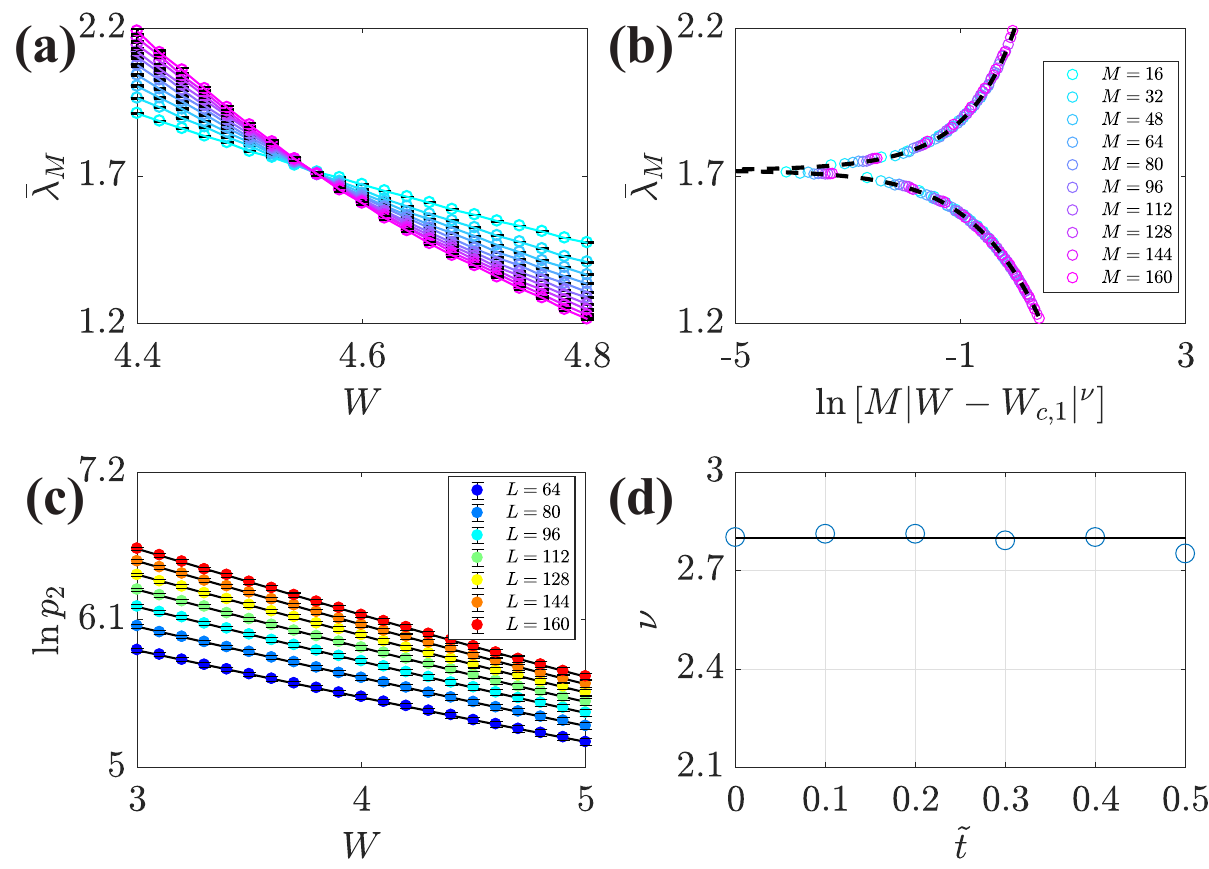}\centering
\caption{(a) \( \bar{\lambda}_M(W) \) for the hybrid-nonreciprocal Rashba model with \( \alpha=0.1 \), \( \tilde{t}=0.1 \), \( E=0 \), and various \( M=16,32,\cdots,160 \). (b) The scaling function \( \bar{\lambda}_M=f(\ln{[M|W-W_{c,1}|^{\nu}]}) \) of data in (a), where the effect of the irrelevant scaling variable is eliminated. The upper and lower branches stand for the boundary-extended and boundary-localized phases, respectively. (c) \( \ln{p_2} \) for various \( L \). The black dashed lines are the scaling function~\eqref{eq_8} obtained through a \( \chi^2 \) fit. (d) Critical exponent \( \nu \) v.s \( \tilde{t} \) of the hybrid-nonreciprocal Rashba model obtained through the scaling analysis of \( \bar{\lambda}_M \) (empty circle) and \( p_2 \) (empty square). The black solid line guides the critical exponent of its reciprocal counterpart. }
\label{fig2}
\end{figure}

The first is through the transfer-matrix method. We perform the calculations on a long strip with a length \(L\geq 10^7\) and a width \(M\sim 100\). The identification of a skin-Anderson transition is similar to that of traditional Anderson transitions: (i) The reduced localization lengths \( \bar{\lambda}_M=\lambda_M/M \) increase (decrease) with \( M \) for boundary-extended (boundary-localized) states; (ii) Near the critical point \( W_{c,1} \), it satisfies a one-parameter scaling law~\cite{cwang_prl_2015}
\begin{equation}
    \begin{gathered}
        \bar{\lambda}_M(W)=f(M/\xi)+\phi M^{-y}
    \end{gathered}\label{eq_8}
\end{equation}
with \( \xi\propto|W-W_{c,1}|^{-\nu} \), \( \phi \) being a constant, and \( y>0 \) being the irrelevant exponent. \( f(x) \) is the universal scaling function. The critical disorder \( W_{c,1} \) and critical exponent \( \nu \) is obtained through a chi-square fit of \( \{ \bar{\lambda}_M \} \) to the scaling function Eq.~\eqref{eq_8}; see Supplementary~\cite{supp}. 
\par

Figure~\ref{fig2}(a) shows a typical example of \( \bar{\lambda}_M \) with \( \alpha=0.1 \), \( \tilde{t}=0.1 \), \( E=0 \), and \( M \) varying from 16 to 160. Through finite-size scaling analysis, we determine the critical disorder \( W_{c,1}=4.554\pm 0.002 \) and \( \nu=2.81\pm0.03 \) for this skin-Anderson transition; see Supplementary~\cite{supp} for more details. Our fit is quite good with a goodness-of-fit \( Q=0.3 \)~\cite{nr_press}, and we plot the smooth scaling function in Fig.~\ref{fig2}(b).  
\par

As an independent check, we further calculate the disorder-average participation ration \( p_2(E)=(\sum_{\bm{i}}|\psi_{E}(\bm{i})|^4 )^{-1} \) with \( \psi_{E}(\bm{i}) \) being the normalized wave function of a right-state of energy \( E=0 \) for the same parameters of Fig.~\ref{fig2}(a). \( p_2 \), obtained through a different approach (the exact diagonalization), also follow a one-parameter scaling near \( W_{c,1} \):~\cite{jhpixley_prl_2015,cWang_prl_2025,cWang_prb_2025}
\begin{equation}
    \begin{gathered}
        p_2(L,W)=L^D\tilde{f}(L/\xi)+\tilde{\phi}L^{-\tilde{y}}
    \end{gathered}\label{eq_9}
\end{equation}
for model~\eqref{eq_1} on a square lattice of size \( L^2 \). In Eq.~\eqref{eq_9}, \( \tilde{f}(x) \) represents the scaling function, \( D \) is the fractal dimension, and the last term \( \tilde{\phi}L^{-\tilde{y}} \) in Eq.~\eqref{eq_8} stands for the scaling of the irrelevant variable. 
\par

\begin{figure}[htbp]
\includegraphics[width=0.48\textwidth]{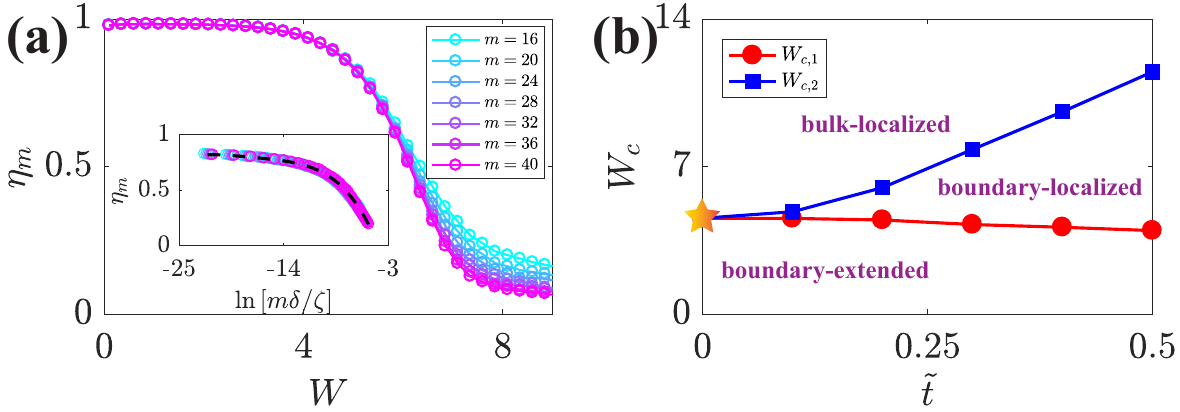}\centering
\caption{(a) \( \eta_m(W) \) of the hybrid-nonreciprocal Rashba model under the same parameters in Fig.~\ref{fig2}(a). Inset: The scaling function near \( W_{c,2} \). (b) A phase diagram of the hybrid-nonreciprocal Rashba model of \( \alpha=0.1 \) and \( E=0 \), determined through the scaling analysis of \( W_{c,1} \) and \( W_{c,2} \), from which one can identify the emergence of boundary-extended, boundary-localized, and bulk-localized phases. For \( \tilde{t}=0 \) (reciprocal limit), \( W_{c,1}=W_{c,2} \), standing for the critical point of the conventional Anderson transition (the pentagram) separating extended and localized states. }
\label{fig3}
\end{figure}

Figure~\ref{fig2}(c) shows \( \ln{p_2} \) as a function of \( W \) using the same parameters as in Fig.~\ref{fig1}(a). We can fit the data using the scaling function from Eq.~\eqref{eq_9}, represented by the black dashed line in Fig.~\ref{fig2}(c). This fitting allows us to obtain \( W_{c,1} = 4.58\pm0.07 \) and \( \nu = 2.73\pm 0.02 \), which align well with the values from \( \bar{\lambda}_M \). Additional fitting parameters and the corresponding scaling function are provided in the Supplementary~\cite{supp}.
\par

The critical exponent \(\nu\) of the hybrid-nonreciprocal Rashba model should be identical to that of its reciprocal counterpart since the transformation from \( T \) to \( \tilde{T} \) does not affect the Lyapunov exponents; see evidence shown in Fig.~\ref{fig2}(d)~\cite{yasada_prl_2002,rSepehrinia_prb_2010,cWang_prb_2017}. In contrast, the fractal dimension at critical points, denoted as \( D \) in Eq.~\eqref{eq_8}, is significantly smaller than that of its reciprocal limit due to the skin effect along the \(x_2\)-direction. However, the fractal dimension \( D \) of the skin-Anderson transition also exhibits universality; see Supplementary~\cite{supp}.
\par

\emph{Boundary-to-bulk localization transition.}$-$As \( W \) increases, a second transition at \( W_{c,2} \) separates boundary- from bulk-localized phases. This transition arises from the competition between two characteristic length scales: the reciprocal localization length \( \xi \) and the skin mode decay length \( \delta \). Specifically, when \( \xi(W)>\delta \), non-Hermitian skin effects dominate, confining states to the boundaries. Conversely, if disorder increases such that \( \xi(W)<\delta \), disorder-induced localization takes over, leading to exponential bulk localization, even with open boundary conditions. The boundary-to-bulk transition point is thus defined by \( \xi(W_{c,2}) = \delta \)~\cite{nHatano_prl_1996,kKawabata_prl_2021}.
\par

Since previous scaling analyses neglect boundary effects and cannot determine \( W_{c,2} \), we instead define \( \eta_m=\sum_{\bm{i}\in\mathcal{D}} |\psi_E({\bm{i}})|^2 \) for system sizes \( m\delta \times 2\delta \) (\( m\in\mathbb{Z},m>2 \)), where \( \mathcal{D} \) is a boundary domain with decay length \( \delta \). Consequently, \( \eta_m \) represents the state's probability weight in \( \mathcal{D} \); it remains constant for the boundary-localized phase but decays with \( m \) for the bulk-localized phase.
\par

Figure~\ref{fig3}(a) displays \( \eta_m \) as a function of \( W \) for various \( m \) for the same parameters as Fig.~\ref{fig2}(a), from which one can see that data of different \( m \) merge and approach to 1 for \( W<W_{c,2} \) and decreases with \( m \) for \( W>W_{c,2} \). This feature indicates the boundary-localized states exist for \( W\in[W_{c,1},W_{c,2}] \), while the bulk-localized states for \( W>W_{c,2} \).
\par

Stimulated by scaling analysis of topological corner to localized state transitions in quantized quadrupole insulators~\cite{cWang_prb_2024}, we propose a scaling law: \( \eta_m(W)=g(m\delta/\zeta) \) for \( W>W_{c,2} \), with \( \zeta \) as the correlation length and \( g(x) \) as a scaling function. We tested two divergence forms for \( \zeta \): a polynomial (\( \zeta\propto |W-W_{c,2}|^{-\nu} \), akin to skin-Anderson transitions) and an exponential (\( \zeta\propto \exp[\alpha/\sqrt{|W-W_{c,2}|}] \), characteristic of Berezinskii-Kosterlitz-Thouless-like transitions)~\cite{vlBerezinskii_JETP_1971,jmKosterlitz_jpc_1973,xcXie_prl_1998,yyZhang_prl_2009}. Our data favor exponential divergence (see the inset in Fig.~\ref{fig3}(a) and the supplementary for details~\cite{supp}). 
\par

We present a comprehensive phase diagram in Fig.~\ref{fig3}(b) for the hybrid-nonreciprocal Rashba model, defined by \( W_{c,1} \) and \( W_{c,2} \) for various \( \tilde{t} \), which delineate three distinct phases in Fig.~\ref{fig1}. Numerical results show \( W_{c,1} \) slightly decreases with \( \tilde{t} \) due to nonreciprocity amplifying randomness. Conversely, \( W_{c,2} \) increases with \( \tilde{t} \), as an amplified non-Hermitian skin effect requires greater disorder to satisfy \( \xi(W_{c,2})=\delta \). We interpret the phase boundaries in Fig.~\ref{fig3}(b) in the Supplementary~\cite{supp}. 
\par

\emph{Discussions.}$-$(i)~The skin-Anderson transition is not restricted to the hybrid-nonreciprocal Rashba model. Any reciprocal system that exhibits an Anderson transition can undergo a skin-Anderson transition once hybrid-nonreciprocity is introduced. We provide two additional examples in the Supplementary~\cite{supp}: a hybrid-nonreciprocal SU(2) model in 2D with effective SU(2) spin-orbit coupling, and a hybrid-nonreciprocal Anderson model in three dimensions. 
\par

(ii)~For simplicity, we focus on a specific kind of hybrid-nonreciprocal systems where reciprocity exists in only one hopping direction and nonreciprocity along the others. However, the skin-Anderson transition actually occurs whenever reciprocal and non-reciprocal hopping terms coexist. Furthermore, the non-reciprocal hopping strengths need not be uniform. However, systems that lack an Anderson transition in the reciprocal limit (such as the two-dimensional Gaussian orthogonal ensemble~\cite{fEvers_rmp_2008}) do not exhibit a skin-Anderson transition and instead show only boundary-to-bulk localization transitions. 
\par

(iii)~Unlike some other anisotropic Anderson transitions predicted in Hermitian systems~\cite{zXiao_prl_2023}, the emergence of nonreciprocity is a crucial factor for observing the skin-Anderson transition. This phenomenon occurs in various real systems, including photonic and microwave lattices, electric circuit networks, cold atoms in optical lattices, and acoustic and mechanical metamaterials. We present one possible scenario for realizing the skin-Anderson transition; further details are provided in the Supplementary materials~\cite{supp}.

\emph{Conclusion.}$-$In conclusion, we predict a disorder-driven skin-Anderson transition in hybrid nonreciprocal systems. This transition occurs at a critical disorder threshold, \( W_{c,1} \), which separates the boundary-extended phase from the boundary-localized phase. The correlation length diverges polynomially as \( \xi \propto |W - W_{c,1}|^{-\nu} \), with the exponent \( \nu \) being consistent with that observed in its reciprocal limit. A second transition occurs at a higher disorder level, \( W_{c,2} \), transitioning from the boundary-localized phase to the bulk-localized phase. Finite-size scaling indicates that this transition resembles a Berezinskii-Kosterlitz-Thouless-like transition, characterized by an exponentially diverging correlation length.
\par

\begin{acknowledgments}

This work is supported by the National Natural Science Foundation of China (Grants No.~12574023).\par

\emph{Data availability.}$-$The raw data supporting the findings of this manuscript are available on Zendo. The custom codes for calculating data are available from the corresponding authors upon reasonable requests.\par 

\end{acknowledgments}

\end{document}